Feasibility of an Infrared Parallax Program Using the Fan Mountain Tinsley Reflector

Jennifer Lynn Bartlett[a1], Chan Park[b2], Srikrishna Kanneganti[c], and Philip A. Ianna[d]

[a]University of Virginia, Department of Astronomy

Post Office Box 400325, University of Virginia, Charlottesville, VA 22904-4325 USA

jennifer.bartlett@usno.navy.mil

[b]University of Virginia, Department of Astronomy

Post Office Box 400325, University of Virginia, Charlottesville, VA 22904-4325 USA

chanpark@kasi.re.kr

[c]University of Virginia, Department of Astronomy

Post Office Box 400325, University of Virginia, Charlottesville, VA 22904-4325 USA

sk4zw@virginia.edu

[d]University of Virginia, Department of Astronomy

Post Office Box 400325, University of Virginia, Charlottesville, VA 22904-4325 USA

p.ianna@juno.com

Corresponding Author:  Jennifer Lynn Bartlett, U.S. Naval Observatory

3450 Massachusetts Avenue, NW, Washington, DC 20392 USA

jennifer.bartlett@usno.navy.mil

001-202-762-1475 Office

001-202-762-1612 FAX

---

[1] Present address:  U.S. Naval Observatory (3450 Massachusetts Avenue, NW, Washington, DC 20392 USA) jennifer.bartlett@usno.navy.mil

[2] Present address:  Korea Astronomy and Space Science Institute (Korea Astronomy and Space Science Institute, Daejeon 306-348, Republic of Korea)  chanpark@kasi.re.kr

ABSTRACT

Despite the continuing importance of ground-based parallax measurements, few active programs remain. Because new members of the solar neighborhood tend towards later spectral types, infrared parallax programs are particularly desirable. Therefore, the astrometric quality of the new infrared camera, FanCam, developed by the Virginia Astronomical Instrumentation Laboratory (VAIL) for the 31-in (0.8-m) Tinsley reflector at Fan Mountain Observatory was assessed using 68 J-band exposures of an open cluster, NGC 2420, over a range of hour angles during 2005. Positions of 16 astrometric evaluation stars were measured and the repeatability of those positions was evaluated using the mean error in a single observation of unit weight. Overall, a precision of $1.3 \pm 0.7$ μm in x (RA) and $1.3 \pm 0.8$ μm in y (Dec) was attained, which corresponds to $0.04" \pm 0.02"$ in each axis. Although greater precision is expected from CCDs in the visual and near-infrared, this instrument can achieve precision similar to that of the ESO NTT infrared parallax program. Therefore, measuring parallaxes in the infrared would be feasible using this equipment. If initiated, such a program could provide essential distances for brown dwarfs and very low mass stars that would contribute significantly to the solar neighborhood census.

1. INTRODUCTION

Even in the well-surveyed northern hemisphere, only 1,131 stellar systems out of an expected 3,125 systems are identified as being within 25 pc (K. Slatten 2009, private communication; Henry et al. 2002). The known systems incorporate the bright nearby stars filled in by the Hipparcos Space Astrometry Mission (ESA 1997, hereafter Hipparcos). Although the Hipparcos catalog includes stars as faint as 12.4 mag in V-band, it is significantly incomplete for stars fainter than 9 mag (Perryman et al. 1997). These limits correspond to an M5V star at approximately 10.5 pc and 2.2 pc, respectively (Drilling & Landolt 2000). Most of the missing systems must, therefore, consist of intrinsically faint stars and brown dwarfs. Such cool objects will have their peak emission at wavelengths of 0.9 μm and longer (Drilling & Landolt 2000). Infrared surveys, such as the Two Micron All Sky Survey (Skrutskie et al. 2006, hereafter 2MASS) and the Deep Near Infrared Survey of the Southern Sky (Epchtein et al. 1999, hereafter DENIS) are fertile sources of potential nearby stars.

Infrared surveys also have identified significant numbers of brown dwarfs, which cool continuously because they lack sufficient mass to sustain nuclear fusion. Trigonometric parallaxes of these substellar objects would allow accurate determination of their luminosities and better knowledge of their velocities. According to Chabrier and Baraffe (2000), most brown dwarf radii should be commensurable with that of Jupiter. Therefore, accurate luminosities would improve temperature estimates in addition to developing more rigorous luminosity functions. More well-determined parameters of brown dwarfs for which parallactic distances can be measured would also enhance our techniques for estimating distances and related properties for those even farther away. Gelino, Kirkpatrick, and Burgasser (2004)[3] list 752 L and T dwarfs of which only 86 have parallax measurements. This list includes the nine relative parallaxes for T dwarfs measured by Tinney, Burgasser, and Kirkpatrick (2003; hereafter TBK03) using the European Southern Observatory New Technology Telescope (ESO NTT) camera and forty preliminary parallaxes for L and T dwarfs measured by Vrba et al. (2004) using the United States Naval Observatory (USNO) 61-inch (1.55-meter) reflector. The Cerro Tololo Inter-American Observatory Parallax Investigation (CTIOPI) currently includes some southern L dwarfs that are thought to lie within the solar neighborhood. However, all brown dwarfs within the reach of ground-based astrometry should have their distances measured.

---

[3]DwarfArchives.org maintains a list of late type stars at www.DwarfArchives.org. Counts reported herein were accurate as of 2009 November 16.

From 2000 until 2006, the USNO operated an infrared parallax program in Flagstaff, Arizona, using an astrometric, infrared camera known as ASTROCAM. In response to the need for brown dwarf parallax measurements, Vrba et al. published preliminary parallaxes for 40 L and T dwarfs in 2004. Unfortunately, a Dewar explosion during a loss of electrical power due to a nearby forest fire in June 2006 severely damaged ASTROCAM. The program is suspended while the instrument is being replaced (F. Vrba 2009, private communication). Even while the USNO program was in full operation, the need for an additional infrared parallax program to increase the number of brown dwarfs with such determinations was clear (Vrba et. al 2004). Assuming one week of observations each month, a parallax program should be able to observe about fourteen sources per hour of right ascension (T. Henry 2006, private communication). Observations of a particular object are usually grouped in semi-annual seasons of high parallax factor and spread over several years in order to separate parallax and proper motion appropriately. Although TBK03 were able to obtain adequate relative parallaxes for eight brown dwarfs in just under two years, three years of observations are more usual. Vrba et al. (2004)

anticipated that three to four years of observations would be necessary before finalizing their larger sample.

The University of Virginia (UVa) has been a leader in the measurement of parallaxes with programs at Leander McCormick, Fan Mountain, and Siding Spring Observatories. However, the last of these, the UVa Southern Parallax Program (Ianna 1993; hereafter SPP) at Siding Spring Observatory, Australia, discontinued observations in July 2002. Once the Virginia Astronomical Instrumentation Laboratory (VAIL) installed an infrared camera, known as FanCam, at Fan Mountain Observatory (S. Kanneganti et. al. 2009; hereafter KC09), this feasibility study of a new parallax program was instituted (Bartlett 2007; Bartlett et al. 2010).

2. INSTRUMENTATION

UVa operates Fan Mountain Observatory (78° 41.6' E, 37° 52.7' N, 556 m elevation) at a dark site in Covesville, Virginia, approximately 19 miles (30 kilometers) south of Charlottesville. Rising 366 meters above the countryside, the isolated peak is above the nighttime inversion layer. The nearest of the Blue Ridge Mountains, which can produce turbulence, is at least 24 kilometers away (Birney 1966). Although seeing conditions better than a second of arc have been reported at this site, recent measurements indicate a median seeing of 1.5" (KC09). In addition to the 31-inch (0.8-meter) reflector discussed herein, Fan Mountain hosts a 40-inch (1-meter) astrometric reflector with spectrograph and 10-inch (0.3-meter) astrograph. Norfolk State University is currently building the 0.6-meter Rapid Response Robotic Telescope (RRRT) there.

Tinsley Laboratories built the 31-inch reflector in the 1960s. Observa-Dome fabricated the 7-meter aluminum dome under which it is mounted. A cinder block building houses the telescope and its control room. Initially, the telescope was used with a photoelectric photometer and photographic-plate spectrograph (Birney 1966). Then, H. J. Wood (2009, private communication) and his students built a two-channel photometer for simultaneous measurements of $H_\beta$ lines in Ap stars with spectrum variability. Also, R. Berg (2009, private communication) used it to design and test a fast photometer to measure stellar diameters from lunar occultations. However, the telescope fell into disuse over the intervening decades. In 2004, VAIL, led by M. Skrutskie, refurbished the telescope and outfitted it with FanCam. Figure 1 shows the facility as it appears today.

VAIL designed and developed FanCam for studying the variability of young stars, measuring fundamental properties for low mass stars and brown dwarfs, observing asteroids and supernovae, and investigating carbon stars identified by 2MASS. Current projects also include observations of blue compact dwarf galaxies and Titan. FanCam saw first light in December 2004 and has been in regular use since January 2005. When mounted on the 31-inch telescope, FanCam has a 0.51" $pixel^{-1}$ (27.56" $mm^{-1}$) scale and an 8.7' field-of-view. A cryogenic refrigeration system cools the detector to 80 K and the filters and optics to 100 K (KC09); these temperatures vary less than 1 K during a night of observing. This "Cryotiger" by IGC Polycold is capable of maintaining the required temperatures for about five weeks during the summer and about five months during the winter (KC09). Tables 1 and 2 describe the telescope and camera further. In addition, the mechanical drawing of FanCam in Figure 2 illustrates some of its significant features.

In a separate analysis, KC09 compared eighty-five stars from NGC 2420 frames with 2MASS images and determined that 51% of the FanCam stellar images were within 50 mas of the 2MASS positions. For stars with $K_S$ magnitudes between 9 and 14, 2MASS positions are internally repeatable to within 40 to 50 mas and show residuals of 70 to 80 mas (Cutri 2005) when compared with the positions in the USNO CCD Astrograph Catalog (Zacharias *et al*. 2000, hereafter UCAC). In addition, KC09 compared the positions of forty-four bright stars in a set of one hundred, 60-second images of NGC 2420 made in J-band. From the scatter in these positions, they estimate the internal repeatability of such measurements to be 1.3 μm (0.035").

FanCam contains two filter wheels with eight positions each. One position in each wheel will remain empty to allow them to operate autonomously. In addition to a blocker and two polarizers, nine filters are installed. The J and H filters share a common focus; the shift between the focus of these two filters and that of the $K_S$ filter is negligible. Table 3 details the J, H, and $K_S$ filters considered for this study; these were salvaged from the 2MASS Prototype Survey camera (Beichman et al. 1998). Images of bright stars taken with this J filter show ghost images that are 5.5 mag fainter than and about 17.5" from the primary image. A ghost-free J filter has replaced the original 2MASS Prototype Survey J filter (KC09).

Methane absorption in T dwarf spectra brings their effective wavelengths nearer to those of typical reference stars when observed through a J or H filter (TBK03). Homogeneous apparent brightnesses for the science target and its associated reference frame, along with the general reduction in differential color refraction (DCR) at longer wavelengths (Stone 1984), increase the scheduling window around an object's transit for parallax observations. Of these filters, J-band has the additional advantage of reduced sky brightness compared to the other bands (KC09). Therefore, a J filter is the most likely choice for a parallax program with FanCam (M. Skrutski 2006, private communication). The USNO infrared parallax program observed L dwarfs in H band and T dwarfs in J band with about the same precision in each filter (Vrba et. al. 2004). Similarly, the ESO NTT program observed T dwarfs in J band (TBK03).

Plans for FanCam and the 31-inch reflector include the development of an autoguider; a prototype of which is in progress. However, the telescope currently tracks the sky reliably for at least 30 seconds. Longer exposures are possible, but image motion will occasionally render one useless. Because the detector saturates at 25,000 analog-to-digital units (ADU), exposures longer than 2 minutes tend not to be practical either (KC09). To observe objects as faint as 17th mag, multiple short exposures are combined. Although appropriate plate constants may be able to account for large shifts or rotations due to the stacking of images, observations for this study used single exposures of 30 seconds or less. FanCam images of at least 10 seconds are limited by Poisson noise from air glow and thermal background so the read noise associated with multiple short exposures is not a significant concern for most of the images used.

3. OBSERVATIONS

On three nights in January, February, and November 2005, FanCam observed open cluster NGC 2420 though the J filter; NGC 2420 is characterized in Table 4. Figure 3 identifies the twenty stars initially selected within this region for evaluation, which are also listed in Table 5. Eventually, four of these would have to be dropped.

A total of sixty-eight exposures of NGC 2420 were selected for this analysis. The January and February observations were 2- or 5-second exposures not initially planned for astrometry; the February frames were also made at large hour angles. The average seeing during these observations was 2". The November observations include frames taken specifically for astrometric assessment and additional frames taken to determine the limiting magnitude of FanCam. These exposure times ranged from 10 to 30 seconds. The average seeing in November was 1.7." Overall, the minimum seeing to pixel scale ratio was 1.8, which indicates the point spread function of the stellar images should be well sampled. In general with a median seeing of

1.5" and a resolution of 0.51" pixel$^{-1}$, FanCam images should be slightly oversampled. IRCAM, an early infrared camera tested by the USNO, had a similar pixel scale and did not suffer from undersampled images (Vrba 2004, 2006).

Custom imaging software, astrix (A. Smith 2004, private communication), read out the Focal Plane Array (FPA) at the beginning and end of each exposure. Each readout requires 0.7 seconds (KC09). The difference between the two readouts was stored as an image in Flexible Image Transport System (FITS) format. Observation times were later obtained from the computer time-stamp associated with the raw frames. Observations were made in batches. The observer moved the telescope slightly while each image was being read out to provide manual dithering. Since then, VAIL developed new software with an enhanced user interface and increased functionality, including some automated dithering.

In November, a series of identical length exposures was made close to the astrometric calibration field for use in estimating the sky background. Previously, the science images themselves were used to estimate the sky background. When the sky was clear near sunset or sunrise, sky flats and dark exposures were also obtained for calibration purposes and used for succeeding nights until a new set could be obtained.

## 4. REDUCTION

During read-out, rows 511 and 512 of each raw image were displaced. Therefore, the first step in the reduction replaced these two rows in each frame with pixels from rows 1,023 and 1,024 shifted by one column to the left using a custom Image Reduction and Analysis Facility (IRAF[4]) routine. The current FPA control software has eliminated this problem. Next, all of the images were trimmed to eliminate bad columns and rows caused by the misalignment of the camera with the focal plane aperture.

---

[4]IRAF is distributed by the NOAO, which are operated by the AURA, under cooperative agreement with the NSF.

Flattened images of NGC 2420 were prepared in batches by night and exposure time. Custom IRAF routines combined the flats, darks, and sky background frames into single representative images. The sky background frames had the same exposure length as the science frames that they would be used to flatten. The representative dark image was subtracted from the representative sky flat to produce a master flat. The representative sky image was subtracted from each science image. Then, the sky-subtracted science images were divided by the master flat image. Finally, previously identified bad pixels were fixed; more than 99% of the array is usable (KC09).

After initial processing, the intensities of some pixels were negative. To correct these intensities, the value of every pixel in a particular image was uniformly increased to bring the minimum above zero.

After this flattening procedure, some visible structure remained in a number of images. Samples of the sky background indicate this variation is significantly less than 1%. This type of artifact is rarely seen in FanCam images and has not been reported since. The flattened images were rotated so that north was to the top and east was to the left. The astrometric evaluation stars were checked individually for saturation and distortion. Next, Starlink conversion routines produced copies of the images in the format used by Figaro Version 2 for further processing.

Begam, Ianna, and Patterson (2010, in preparation) describe the Figaro-based image reduction pipeline developed for the SPP. Following these procedures, the images were cataloged and graded. The astcor routine extracted the position of each astrometric evaluation star. In some frames where large constants had been added to ensure all pixels contained positive intensities, the constants had to be subtracted again so that astcor did not misidentify bright stars as saturated.

Once positions for individual stars were measured, several combinations of images were considered: all images, individual nights, January and November images with moderate hour angles, and February and November images with large hour angles. Parallax input files were constructed for each case treating the central star, astrometric evaluation star 20, as the "parallax" star. For individual nights and for the February and November combination, the image with the best quality overall along with the image taken immediately after it were averaged to serve as the "trail plate," to which system the other frames were reduced. For the other combinations, four January images taken close to the meridian were averaged as the trail plate. Averaging several images increases the coordinate precision associated with the trail plate. The January trail plate lacked astrometric evaluation stars 1, 2, 3, and 10 so those stars were dropped from further analysis.

The McCormick Parallax Reduction Program (MPRP) reduced the positions of the astrometric evaluation stars in each additional image to the selected trail plate. MPRP calculated a three plate-constant model to describe the mean positions of the astrometric evaluation stars; the plate constants accounted for scale, orientation, and origin.

The mean error in a single observation of unit weight (m.e.1), estimated herein using the standard deviation of the plate adjustment model for each frame, is a measure of repeatability of positions (Monet et al. 1992). Table 6 lists the averages for each batch of frames considered. The x- and y-coordinates, which correspond to right ascension and declination, are considered separately. Overall, the precision is 1.3 ± 0.7 μm and 1.3 ± 0.8 μm in the x- and y-axes, respectively, which corresponds to 0.04" ± 0.02" in either case. The errors in a particular axis range from 0.6 to 2.0 μm (0.016 to 0.06"). The early November frames that were taken specifically as astrometric calibration frames have the smallest errors in x and y. This batch has the advantage of including only background-limited images collected within 15 minutes of the meridian under better seeing conditions than occurred earlier in the year. The batches taken at large hour angles (February, November late, and both combined) have larger errors in y than x, as may be expected. The January frames taken within five minutes of the meridian have the largest errors, with its x-coordinate slightly worse than the y, possibly due to shorter exposure times and poorer seeing.

5 ANALYSIS

The xy-averaged precision for all frames considered is about ±1.3 μm (±0.04"), which is slightly better than the single-position precision of ±1.7 μm (Guinan & Ianna 1983) achieved for photographic plates. The average m.e.1 for the SPP CCD #6 is ±0.39 (±10 mas) with Cousins V, R, and I filters (Bartlett, Ianna, & Begam 2009) while the USNO optical parallax program obtains about ±0.2 μm (3 mas) with filters similar to R, I, and $z$ (Harris et. al 2005). The November images made by FanCam within 15 minutes of the meridian come closest to these optical CCD programs. While the FanCam images of NGC 2420 did not quite achieve the precision of these CCD programs, photographic plates were the detector of choice for parallax programs throughout most of the twentieth century. Therefore, a parallax program should be possible using FanCam as well.

More apropos comparisons would be those of FanCam to other infrared parallax programs: the current infrared parallax program at the USNO and the brief program at the ESO NTT. Table 7 characterizes these detectors while Table 8 summarizes their precision. First, USNO preliminary astrometric testing using cluster fields obtained a total m.e.1 of ±0.5 μm (±7 mas) in J-band using IRCAM, which was not specifically designed for astrometry (Vrba et al. 2000, 2004). The total m.e.1 for FanCam is 1.9 μm (±50 mas) overall or ±1.0 μm (±30 mas) for the November frames alone. Encouraged by their IRCAM results, the USNO then developed an infrared detector expressly for astrometry, ASTROCAM. Vrba (2009, private communication) reports an average m.e.1 for ASTROCAM of about 0.26 μm (3.5 mas). In addition to the USNO infrared parallax program, TBK03 measured relative parallaxes for nine T dwarfs in J band using the Son of ISAAC[5] (SofI) infrared camera on the ESO NTT. They report a median m.e.1 of 0.78 μm (12.1 mas) for their program. The median m.e.1 for FanCam overall is 1.1 μm (30 mas) or 0.6 μm (20 mas) for only the November frames. While the overall precision of FanCam measured here is not as good as either of these infrared parallax programs, the November FanCam frames reach the precision of the ESO NTT program in their respective image planes.

The lower precision of FanCam does not preclude it from contributing valuable parallaxes to the nearby star census. Actual parallaxes measured over an extended period of time and along the full parallactic ellipse should improve roughly with the square root of the number of observing nights, assuming reasonable conditions and an adequate reference frame. Parallaxes with errors less than 10 mas (Vrba 2009, private communication), which correspond to a 10% error in distance at 10 pc, should be attainable over 4 years or so.

6. DISCUSSION

Although the overall m.e.1 of FanCam positions measured for the astrometric evaluation stars selected in NGC 2420 are not quite as good as results achieved elsewhere using CCDs and FPAs in the visual and infrared bands, the November FanCam frames matched the precision of the ESO NTT program. Furthermore, the FanCam data are slightly better in quality to those obtained from photographic plates. Therefore, the measurement of parallaxes using FanCam is possible. However, additional testing would be worthwhile before committing substantial amounts of observing time to a new infrared parallax program. Such testing should include the

---

[5] ISAAC is the Infrared Spectrometer And Array Camera.

evaluation of possible higher-order plate constants, the analysis of stacked images, the assessment of other proposed filters, and the measurement of an actual parallax. Additional astrometric improvements could be obtained by

- increasing the number of exposures used, which should not be difficult if background-limited exposures shorter than 30 seconds are used.
- refining the initial image reduction process to obtain flatter images or eliminating any future images with residual structure.
- optimizing the parallax reduction pipeline for use with infrared images and with the new hardware, which should avoid the need for different approaches to background subtraction.
- installing the planned autoguider, development of which is in progress.
- modeling the atmospheric refraction so that DCR corrections may be included as necessary; although the effect should be small in the infrared, DCR corrections may improve measurements made at large hour angle.
- establishing consistent astrometric observing procedures, including guidelines for grading frames.

Along with the upgrades already accomplished, such systematic enhancements should increase the precision, ameliorate some awkward features of the current system, and make a parallax program more practical. These FanCam results also indicate that a similar investigation of Peters Automated Infrared Imaging Telescope (PAIRITEL) on Mount Hopkins may also be worthwhile. While PAIRITEL would have only limited time for astrometry, it is capable of detecting fainter objects than the smaller FanCam.

Although the USNO infrared parallax program may return in early 2010 (F. Vrba 2009, private communication), another such program in the northern hemisphere would still be beneficial. In this region alone, a substantial number of objects remains to be studied with more than 60% of the expected stellar systems within 25 pc still undetected and nearly 90% of brown dwarfs without reliable distances. At 25 pc, an M5V star will have a J-band magnitude of 8 (Drilling & Landolt 2000; Tokunaga 2000), which is much brighter than the limiting J-magnitude of 19 for FanCam (KC09). Similarly, L5 and T5 dwarfs would be about 15 and 16 magnitude (Gelino, Kirkpatrick, & Burgasser 2004). About 119 northern hemisphere brown dwarfs are possible members of the solar neighborhood from which an initial target list could be developed (K. Slatten 2009, private communication). In addition to increasing the number of cool objects for which trigonometric parallaxes are available, some overlap between the programs will allow the inter-comparison of results. FanCam is an opportunity to make substantial contributions to our understanding of the solar neighborhood and of substellar bodies.

## 7. ACKNOWLEDGEMENTS

This research used the NASA ADS Bibliographic Services; Aladin; VizieR catalog access tool and SIMBAD database, operated at CDS, Strasbourg, France; the WEBDA database, operated at the Institute for Astronomy of the University of Vienna; data products from 2MASS, which is a joint project of the University of Massachusetts and IPAC/CalTech, funded by NASA and NSF; and the M, L, and T dwarf compendium housed at DwarfArchives.org and maintained by C. Gelino, D. Kirkpatrick, and A. Burgasser. In addition, we acknowledge the data analysis facilities provided by the Starlink Project, which is run by the CCLRC on behalf of PPARC.

M. Skrutskie generously made observing time and archival observations available. M. Begam, who developed the Figaro routines and pipeline, thoughtfully provided advice and additional support with the data reduction and analysis. R. Berg, D. Birney, L. Frederick, and H. J. Wood cordially reminisced about the telescope during the 1960s. F. Vrba graciously shared his experience with the USNO infrared parallax program.

K. Slatten maintains an up-to-date list of the nearest stars in association with the Nearby Star Observers. He enthusiastically provided the counts of known systems and brown dwarfs within 5, 10, and 25 parsecs used herein.

K. Riggleman kindly shared the photographs of 31-inch Tinsley reflector that he took in November 2009. John Bartlett helpfully processed those pictures for inclusion in this article.

The F. H. Levinson Fund of the Peninsula Community Foundation, UVa Governor's Fellowship and Graduate School of Arts and Sciences, Hampden-Sydney College, and the USNO funded this research.

8. REFERENCES

Bartlett, J. L. 2007, "Knowing Our Neighbors: Fundamental Properties of Nearby Stars," Ph.D. diss., UVa

Bartlett, J. L., Ianna, P. A., Begam, M. C. 2009, PASP, 121, 365

Bartlett, J. L., Park, C. Kanneganti, S. & Ianna, P. A. 2010, Bull. AAS, 41, 401

Beichman, C. A., Chester, T. J., Skrutskie, M., Low, F. J., & Gillett, F. 1998, PASP, 110, 480

Beletic, J. W. et al. 2008, in Proc. SPIE 7021, High Energy, Optical, and Infrared Detectors for Astronomy III, ed. D. A. Dorn & A. D. Holland (Bellingham, WA: SPIE) 70210H

Birney, D. S. 1966, S&T, 31, 210

Cannon, R. D. & Lloyd, C. 1970, Mon. Not. R. Astron. Soc., 150, 279

Chabrier, G. & Baraffe, I. 2000, Annu. Rev. Astron. Astrophys., 38, 337

Cutri, R. M. 2005, in Explanatory Suppl. to the 2MASS All Sky Data Release, R. M. Cutri et al. (Pasadena, CA: IPAC)

http://www.ipac.caltech.edu/2mass/releases/allsky/doc/sec2_2.html#pscastrprop

Drilling, J. S. & Landolt, A. U. 2000 in Allen's Astrophysical Quantities, ed. A. N. Cox (4th ed.; New York: Springer-Verlag)

Epchtein, N. et al. 1999, Astron. Astrophys., 349, 631 (DENIS)

ESA. 1997, The Hipparcos and Tycho Catalogues, (ESA SP-1200) (Noordwijk, Netherlands: ESA Publ. Div.) (Hipparcos)

Farris, M. 2004, Rockwell Scientific FPA Test Data Summary (Camarillo, CA: Rockwell Scientific)

Friel, E. D., Janes, K. A., Tavarez, M., Scott, J., Katsanis, R., Lotz, J., Hong, L., & Miller, N. 2002, Astron. J., 124, 2693

Gelino, C. R., Kirkpatrick, J. D., & Burgasser, A. J. 2004, Bull. AAS, 36, 1354

Guinan, E. F. & Ianna, P. A. 1983, AJ, 88, 126

Harris, H. et al. 2005, in ASP Conf. Ser. 338, Astrometry in the Age of the Next Generation of Large Telescopes, ed. P. K. Seidelmann & A. K. B. Monet (San Francisco: ASP), 122

Henry, T., Walkowicz, L. M., Barto, T. C., & Golimowski, D. 2002, Astron. J., 123, 2002

Høg, E. et al. 2000, Astron. Astrophys., 355, L27

Ianna, P. A. 1993, in IAU Symp. 156, Developments in astrometry and their impact on astrophysics and geodynamics, ed. I. Mueller & B. Kolaczek (Dordrecht: Kluwer Academic Publ.), 75 (SPP)

Kanneganti, S., Park, C., Skrutskie, M. F., Wilson, J. C., Nelson, M. J., Smith, A. W., & Lam, C. R. 2009, PASP, 121, 885 (KC09)

Loktin, A. V. & Beshenov, G. V. 2003, Astron. Rep., 47, 6

Monet, D. G., Dahn, C. C., Vrba, F. J., Harris, H. C., Pier, J. R., Luginbuhl, C. B., & Ables, H. D. 1992, Astron. J., 103, 638

Monet, D. et al. 2003, Astron. J., 125, 984 (USNO-B1.0)

Perryman, M. A. C. et al. 1997, Astron. Astrophys., 323, L49

Skrutskie, M. F. et al. 2006, Astron. J., 131, 1163 (2MASS)

Stone, R. C. 1984, A&A, 138, 275

Tinney, C. G., Burgasser, A. J., & Kirkpatrick, J. D. 2003, AJ, 126, 975 (TBK03)

Tinsley Lab. 1963, Optical Layout, D48-1 (Berkeley, CA: Tinsley)

Tadross, A. L. 2001, New Astron., 7, 293

Tokunaga, A. T. 2000 in Allen's Astrophysical Quantities, ed. A. N. Cox (4th ed.; New York: Springer-Verlag)

Vrba, F. 2006, in USNO Astrometry Forum Proc., ed. P. Shankland, (Washington, DC: USNO) submitted

Vrba, F. J., Henden, A. A., Luginbuhl, C. B., Guetter, H. H., & Monet, D. G. 2000, BAAS, 32, 678

Vrba, F. J. et al. 2004, Astron. J., 127, 2948

Xin, Y. & Deng, L. 2005, Astrophys. J., 619, 824

Zacharias, N. et al. 2000, Astron. J., 120, 2131 (UCAC)

TABLES

TABLE 1
CHARACTERISTICS OF THE TINSLEY REFLECTOR AT FAN MOUNTAIN

| Parameter | | Description |
|---|---|---|
| Objective Size | (m) | 0.8 |
| | (in) | 31 |
| Optics | | classical Cassegrain |
| Telescope Focal Ratio | | f/15.484 |
| Telescope Effective Focal Length | (m) | 12.19 |
| | (in) | 480.0 |
| Telescope Focal Plane Scale | (as mm$^{-1}$) | 16.92 |

REFERENCE.—Tinsley Laboratories, 1963

TABLE 2
CHARACTERISTICS OF FANCAM

| Parameter | Description | Reference |
|---|---|---|
| Focal Plane Array | Rockwell Scientific HAWAII-1 HgCdTe | 1, 2 |
| Focal Plane Array Description | thick, back-illuminated, coated | |
| FanCam Field of View (arcmin$^2$) | 8.7 | 1 |
| Detector Size (pixels) | 1,024 x 1,024 | 2 |
| Detector Pixel Size (μm$^{-2}$) | 18.5 | 2 |
| FanCam resolution ("pixel$^{-1}$) | 0.51 | 1 |
| (" mm$^{-1}$) | 27.56 | 1 |
| Spectral Response (μm) | 0.85–2.5 | |
| Field Curvature (μm) | <50 | 1 |
| Distortion, 90% of field (pixel) | <0.1 | 1 |
| at corners (pixel) | 0.5 | 1 |
| Detector Gain (e$^-$ DN$^{-1}$) | 4.6 | 1 |
| Read Noise (e$^-$ rms) | 17 | 1 |
| Dark Current (e$^-$ s$^{-1}$) | <0.1 | 3 |
| Well Capacity (ke$^-$) | >97 | 2 |
| Quantum Efficiency, K (%) | 58.10 | 2 |
| Operating Temperature (K) | 80 | 1, 2 |

REFERENCES.—(1) KC09; (2) Farris 2004; (3) Beletic et al. 2008
NOTES.—Rockwell Scientific is now Teledyne Imaging Systems. HAWAII stands for HgCdTe Astronomical Wide Area Infrared Imager.

TABLE 3
CHARACTERISTICS OF SELECTED FILTERS

| Filter[a] | Source | Passband (μm) | Central Wavelength (μm) | Transmittance (% Average) | Limiting Magnitude[b] |
|---|---|---|---|---|---|
| J[c] | ThrillCam | 1.117 – 1.422 | 1.270 | 85 | 19.0 |
| J[d] | 2MASS Prototype[e] | 1.105 – 1.395 | 1.250 | 75 | 19.0 |
| H | 2MASS Prototype[e] | 1.515 – 1.817 | 1.666 | 92 | 18.0 |
| K$_S$ | 2MASS Prototype[e] | 2.045 – 2.335 | 2.190 | 78 | 17.0 |

NOTES.—[a]Additional filters currently include K, Y, H$_2$, Br-γ, FE-II, Pa-β, and two polarizers.
[b]Limiting magnitude is based on a 10-σ detection after 10 min. of on-source observation (KC09).
[c]This ghost-free filter is currently available for observations.
[d]This filter, which was used in this study, produces ghost images that are 5.5 mag fainter than and 17.5" from bright sources. It has been replaced by a ghost-free filter.
[e]Filters were salvaged from the 2MASS Prototype Survey camera (Beichman et al. 1998).

TABLE 4
PROPERTIES OF NGC 2420

| Parameter | | Value | Reference |
|---|---|---|---|
| Position (2000.0) | RA | 07$^h$38$^m$23$^s$ | 1 |
| | Declination | +21°34'24" | |
| Classification | | Open Cluster | |
| Distance (kpc) | | 3 | 2 |
| Proper Motion (mas yr$^{-1}$) | | 4.3 ± 0.3 | 3 |
| Position Angle of Proper Motion (degrees) | | 198 ± 5 | 3 |
| Radial Velocity (km s$^{-1}$) | | 67 ± 8 | 4 |
| Age (Gyr) | | 2.00 | 2 |
| Metallicity ([Fe/H]) | | -0.38 ± 0.07 | 4 |

REFERENCES.—(1) Xin & Deng 2005; (2) Tadross 2001; (3) Loktin & Beshenov 2003; (4) Friel *et al.* 2002

TABLE 5
ASTROMETRIC EVALUATION STARS SELECTED IN NGC 2420

| Evaluation Star | 2MASS Designation[a] | 2MASS Photometry (mag) J | | | 2MASS Photometry (mag) H | | | 2MASS Photometry (mag) $K_S$ | | | USNO-B1.0 Designation | NGC 2420 ID |
|---|---|---|---|---|---|---|---|---|---|---|---|---|
| 1[b] | 07380627+2136542 | 10.781 | ± | 0.021 | 10.198 | ± | 0.019 | 10.125 | ± | 0.018 | 1116-0164806 | 41 |
| 2[b] | 07381549+2138015 | 10.903 | ± | 0.021 | 10.405 | ± | 0.023 | 10.305 | ± | 0.018 | 1116-0164869 | 76 |
| 3[b] | 07382696+2138244 | 10.590 | ± | 0.021 | 10.057 | ± | 0.022 | 9.982 | ± | 0.018 | 1116-0164970 | 174 |
| 4 | 07382208+2136432 | 10.806 | ± | 0.022 | 10.277 | ± | 0.024 | 10.192 | ± | 0.020 | 1116-0164929 | 119 |
| 5 | 07382195+2135508 | 10.840 | ± | 0.024 | 10.350 | ± | 0.026 | 10.210 | ± | 0.020 | 1115-0161842 | 118 |
| 6 | 07382687+2135460 | 11.987 | ± | 0.022 | 11.904 | ± | 0.022 | 11.861 | ± | 0.020 | 1115-0161940 | 172 |
| 7[c] | 07381507+2134589 | 8.572 | ± | 0.021 | 7.854 | ± | 0.016 | 7.687 | ± | 0.027 | 1115-0161730 | 73 |
| 8 | 07382148+2135050 | 11.345 | ± | 0.022 | 10.805 | ± | 0.024 | 10.707 | ± | 0.018 | 1115-0161830 | 114 |
| 9 | 07382984+2134509 | 11.107 | ± | 0.022 | 10.592 | ± | 0.022 | 10.475 | ± | 0.017 | 1115-0161983 | 192 |
| 10[b] | 07383760+2134119 | 10.848 | ± | 0.022 | 10.358 | ± | 0.024 | 10.234 | ± | 0.020 | 1115-0162073 | 236 |
| 11 | 07382936+2134309 | 12.049 | ± | 0.028 | 12.038 | ± | 0.030 | 11.946 | ± | 0.023 | 1115-0161978 | 190 |
| 12 | 07382166+2133514 | 9.5430 | ± | 0.024 | 8.954 | ± | 0.023 | 8.799 | ± | 0.018 | 1115-0161834 | 115 |
| 13 | 07382696+2133313 | 10.182 | ± | 0.022 | 9.772 | ± | 0.022 | 9.662 | ± | 0.017 | 1115-0161945 | 173 |
| 14 | 07382724+2133166 | 11.792 | ± | 0.022 | 11.331 | ± | 0.022 | 11.216 | ± | 0.018 | 1115-0161953 | 176 |
| 15 | 07382418+2132540 | 9.388 | ± | 0.021 | 8.721 | ± | 0.018 | 8.572 | ± | 0.018 | 1115-0161886 | 140 |
| 16 | 07381822+2132062 | 10.882 | ± | 0.022 | 10.413 | ± | 0.024 | 10.272 | ± | 0.018 | 1115-0161775 | 91 |
| 17 | 07382406+2132148 | 10.758 | ± | 0.029 | 10.279 | ± | 0.031 | 10.180 | ± | 0.026 | 1115-0161881 | 139 |
| 18 | 07382935+2132375 | 11.784 | ± | 0.021 | 11.265 | ± | 0.022 | 11.183 | ± | 0.020 | 1115-0161977 | 188 |
| 19 | 07382114+2131418 | 10.988 | ± | 0.022 | 10.524 | ± | 0.024 | 10.413 | ± | 0.017 | 1115-0161826 | 111 |
| 20[d] | 07382285+2134069 | 12.021 | ± | 0.021 | 11.490 | ± | 0.022 | 11.427 | ± | 0.017 | 1115-0161858 | 126 |

NOTES.—[a]2MASS designations are based on right ascension and declination (J2000) and have the form hhmmss.ss+ddmmss.s.
    [b]Astrometric evaluation stars 1, 2, 3, and 10 were eventually dropped because they did not appear in the January frames averaged for the "trail plate."
    [c]Astrometric evaluation star 7 is also TYC 1373-01207-1 (Høg et al. 2000). It saturated in some of the November frames and was dropped from the analysis of some batches.

[d] The reduction software treated astrometric evaluation star 20 as the expected parallax star. This star was also used to grade frames and estimate seeing.

REFERENCES.—USNO-B1.0 is Monet et al. (2003); NGC 2420 identification follows Cannon & Lloyd (1970).

TABLE 6
AVERAGE STANDARD DEVIATIONS FOR INDIVIDUAL FRAMES COMPARED TO TRAIL PLATES

| Frame Batch | X Coordinate[a] (μm) | Y Coordinate[a] (μm) | XY Averaged[a] (μm) | Comment |
|---|---|---|---|---|
| All | 1.31 ± 0.67 | 1.34 ± 0.77 | 1.32 ± 0.72 | |
| Jan, Nov Early | 1.27 ± 0.80 | 1.24 ± 0.79 | 1.26 ± 0.79 | Frames within 15 minutes of meridian |
| Feb, Nov Late[b] | 0.73 ± 0.34 | 1.33 ± 0.43 | 1.03 ± 0.48 | Frames greater than 50 minutes west |
| Jan Only | 2.02 ± 0.88 | 1.88 ± 0.88 | 1.95 ± 0.75 | 2- and 5-s exposures |
| Feb Only | 1.02 ± 0.57 | 1.73 ± 0.77 | 1.38 ± 0.75 | Frames greater than 160 minutes west, 5-s exposures |
| Nov Only[b] | 0.58 ± 0.58 | 0.83 ± 0.39 | 0.71 ± 0.32 | All frames for November |
| Nov Early[b] | 0.57 ± 0.15 | 0.57 ± 0.15 | 0.69 ± 0.33 | Frames within 15 minutes of meridian |
| Nov Late[b] | 0.60 ± 0.16 | 1.03 ± 0.22 | 0.82 ± 0.29 | Frames greater than 50 minutes west |

NOTE.— [a] The FanCam focal plane scale is 27.56 mas μm$^{-1}$.
[b] Astrometric evaluation star 7 was dropped from these batches because it saturated in the November frames averaged for the "trail plate."

TABLE 7
COMPARISON OF PIXEL SCALES AMONG INFRARED INSTRUMENTS

| Detector | Pixel Size ($\mu m^{-2}$) | Pixel Scale (" pixel$^{-1}$) | Reference |
|---|---|---|---|
| FanCam | 18.5 | 0.51 | 1,2 |
| IRCAM | 40 | 0.54 | 3 |
| ASTROCAM | 27 | 0.3654 | 4 |
| ESO NTT SofI | 18.5 | 0.28826 | 5,6 |

REFERENCES.—(1) Farris 2004; (2) KC09; (3) Vrba 2006; (4) Vrba et al. 2004; (5) ESO 2008;[6] (6) TBK03

TABLE 8
COMPARISON OF ASTROMETRIC PRECISION AMONG INFRARED PARALLAX PROGRAMS

| Program | Average m.e.1 ($\mu m$) | Average m.e.1 (mas) | Median m.e.1 ($\mu m$) | Median m.e.1 (mas) | Reference |
|---|---|---|---|---|---|
| UVa FanCam, all | 1.3 | 40 | 1.1 | 30 | |
| UVa FanCam, Nov only | 0.7 | 19 | 0.6 | 16 | |
| USNO ASTROCAM | 0.2 | 3 | … | … | 1 |
| ESO NTT SofI | … | … | 0.78 | 12.1 | 2 |

NOTE.—TBK03 provides the median m.e.1 rather than average; both FanCam values are provided for comparison.
REFERENCES.—(1) Vrba 2006; (2) TBK03

---

[6]ESO provides SofI detector specifications on-line at
http://www.eso.org/sci/facilities/lasilla/instruments/sofi/inst/HawaiiDetector.html

FIGURES

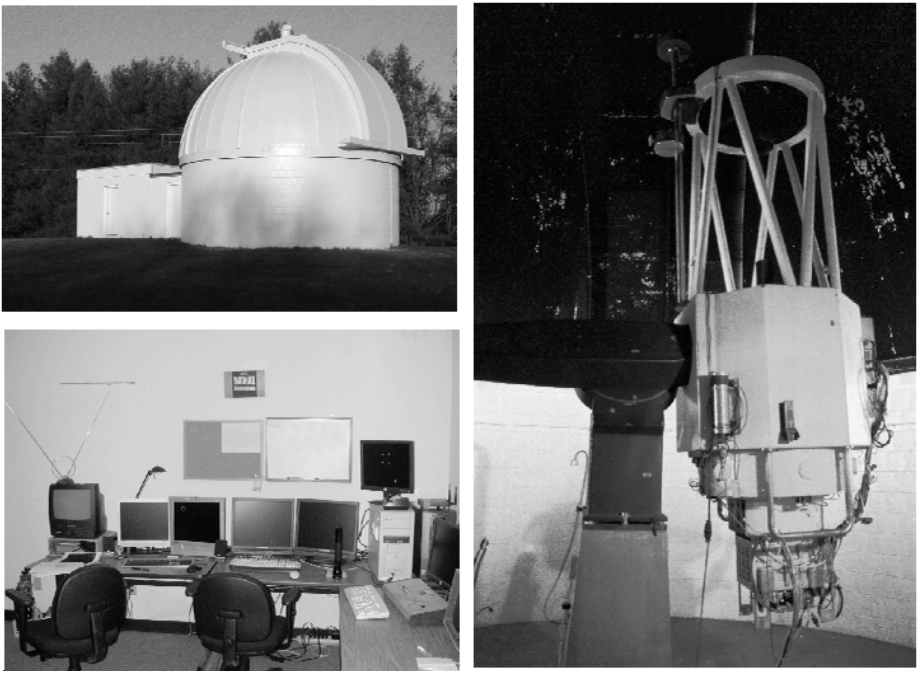

FIG. 1. 31-inch Tinsley Reflector Facility at Fan Mountain Observatory. The upper left image shows the outside of the control room and dome. The right image depicts the telescope itself. The lower left image presents the inside of the control room. (Image credit: K Riggleman & J. Bartlett, Nov. 2009).

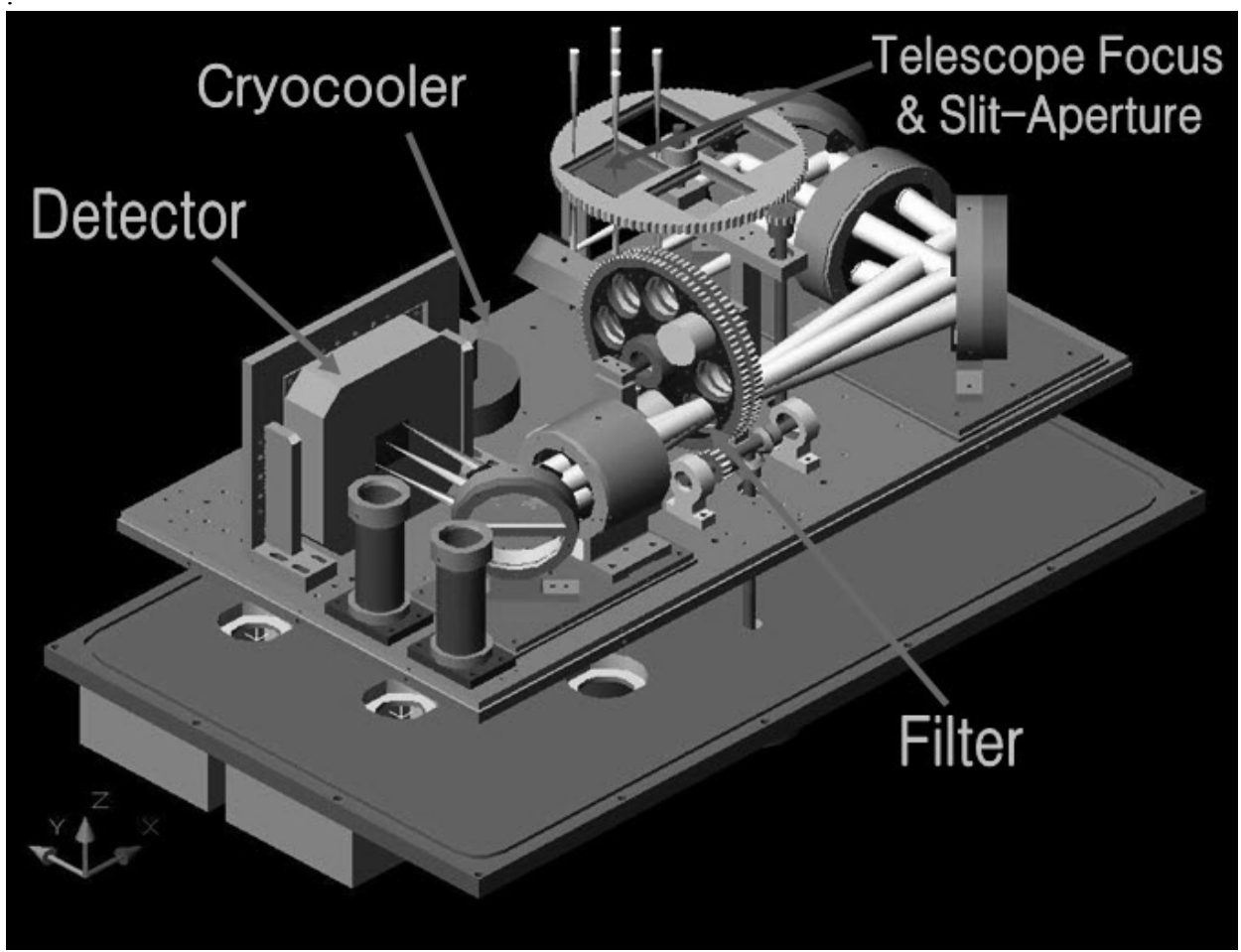

FIG. 2. — FanCam Mechanical Drawing.

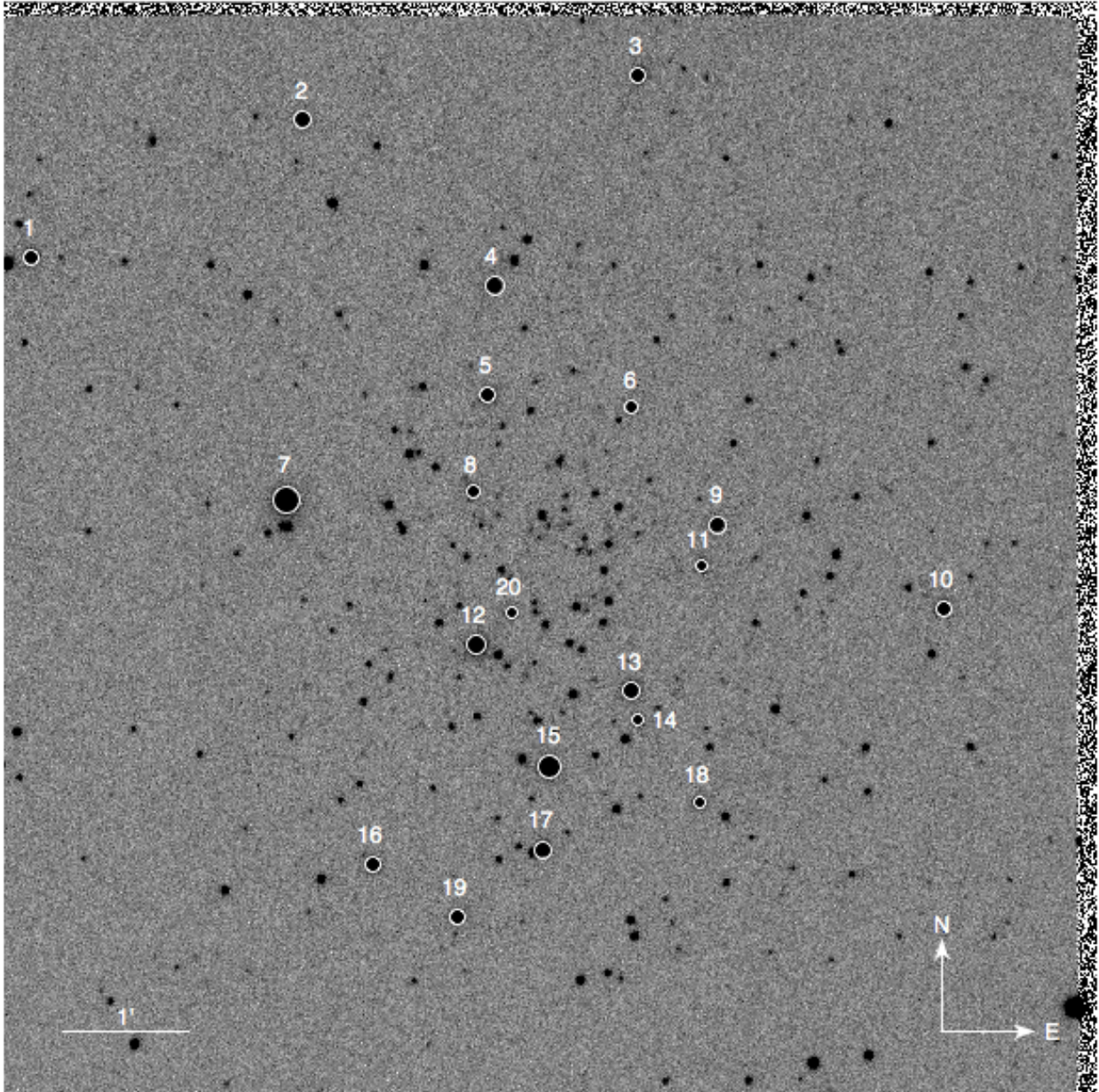

FIG. 3. — NGC 2420 with Astrometric Evaluation Stars Identified. FanCam 10-s, J-band frame taken 2005 May 9. Misalignment of the camera with the focal plane aperture is apparent along the north and east edges. The reduction software treated star 20 as the expected parallax star, which was also used to grade frames and estimate seeing. Stars 1, 2, 3, and 10 were eventually dropped from the analysis because they did not appear in the January frames used as a "trail plate." Star 7 was dropped from several batches because it was saturated in some November frames.